\documentclass[superscriptaddress,prl,aps,twocolumn]{revtex4}
\usepackage{graphicx}
\begin{document}


\title{
Photoemission study of the spin-density wave state in thin films of Cr}

\author{F. Schiller}
\affiliation{Institut f\"ur Festk\"orperphysik, TU Dresden, D-01062 Dresden, Germany}

\author{D.V. Vyalikh}
\affiliation{Institut f\"ur Festk\"orperphysik, TU Dresden, D-01062 Dresden, Germany}

\author{V.D.P. Servedio}
\affiliation{Sezione INFM and Dip.\ di Fisica, Universit\`a ``La Sapienza'', P.le A.\ Moro 2, 
I-00185 Roma, Italy}

\author{S.L. Molodtsov}
\affiliation{Institut f\"ur Festk\"orperphysik, TU Dresden, D-01062 Dresden, Germany}

\date{\today}

\begin{abstract}

Angle-resolved photoemission (PE) was used to characterize the spin-density wave (SDW) state in 
thin films of Cr grown on W(110). The PE data were analysed using results of local spin density 
approximation layer-Korringa-Kohn-Rostoker calculations. It is shown that the incommensurate SDW 
can be monitored and important parameters of SDW-related interactions, such as coupling strength 
and energy of collective magnetic excitations, can be determined from the dispersion of the 
renormalized electronic bands close to the Fermi energy. The developed approach can readily be 
applied to other SDW systems including magnetic multilayer structures.
\end{abstract}


\maketitle

Bulk Cr is an almost unique material revealing itinerant antiferromagnetism with a spin-density 
wave (SDW) ground state at room temperature \cite{Fawc}. At the N{\'e}el temperature T$_N$ = 311 K 
chromium exhibits a transition from a paramagnetic ($bcc$ lattice) to an antiferromagnetic (AF) 
order ($sc$ of CsCl type) that is modulated by the incommensurate SDW along the $<$100$>$ 
directions \cite{Fawc,Trans,Falic}. Thereby the periods of the AF arrangement and the SDW 
oscillations amount to 2 and $\sim$ 21 monolayers (ML) of Cr, respectively. It is widely accepted 
that the AF order (often referred to as the commensurate SDW) is caused by a nesting of the Fermi 
surface (FS) sheets around the $\Gamma$ and the H points of the Brillouin zone (BZ) of $bcc$ Cr, 
while the nature of the incommensurate SDW is still the subject of extensive debates 
\cite{Schilf}. The SDW in Cr is accompanied by a strain wave and a charge-density wave with half 
the period of the SDW \cite{Rieder} as well as by a series of collective excitations including 
spin waves (magnons) and phonons \cite{Fawc}. Electron interactions particularly with the magnetic 
excitations lead to renormalization of the electronic structure of the ground state. Although a 
number of attempts was made to study the renormalization of the electronic bands in some Cr 
systems \cite{Fawc,Dod} the subject requires further investigations. 
  
A detailed understanding of the SDW and the SDW-related phenomena in Cr is of high importance, 
since the above AF short- and SDW long-range magnetic modulations give strong reason to use Cr as 
spacers in magnetic multilayer structures providing giant magnetoresistance, spin-valve effect and 
applications in magnetic sensor technology \cite{Ba}. One of the mostly investigated up to now 
system Fe/Cr/Fe(100) shows that the ferromagnetic or AF type of coupling between Fe layers varies 
with thickness of Cr spacer following the short period, whereas the strength of the coupling 
changes with the long period of oscillations \cite{Par,Pur,Walk,Urg}. The description of the SDW 
in thin films is complicated by the fact that the boundary conditions at the interfaces have to be 
properly considered. In the density-functional theory (DFT) study of Fe/Cr/Fe(100) by Niklasson 
{\it et al.} \cite{Nikl} mainly AF order was found for Cr spacers with thicknesses $<$ 10 ML. For 
thicker layers, various branches of sometimes coexisting SDWs, which differ from each other by the 
number of nodes $m$, were calculated. Upon increase of the Cr thickness, each $m$ branch is 
abruptly substituted by a $(m+2)$ branch giving rise to phase slips of the short-range 
oscillations \cite{Urg}, which, however, may also be correlated with the incommensurability of the 
bulk nesting condition. A similar study within the DFT approach was performed both for 
Mo/Cr/Mo(100) structure and for Cr films on a Mo(100) surface \cite{Nikl2}. The SDW order in 
Fe/Cr/Fe was also treated by means of the Korringa-Kohn-Rostoker Green's function method within 
the framework of the local spin-density functional formalism \cite{Hir}. 
    
While the SDW in Cr films seems to be relatively well investigated theoretically, experimental 
studies are mainly restricted to rather indirect information obtained from measurements of induced 
magnetic properties of marginal layers, which were performed, e.g., by means of the magneto-optic 
Kerr effect \cite{Pur}, spin-polarized electron-energy loss spectroscopy \cite{Walk} and scanning 
electron microscopy with polarization analysis \cite{Urg}. So far, no systematic photoemission 
(PE) study of the SDW in Cr films of different thicknesses, except the work of Sch\"afer {\it et 
al.} \cite{Schaf} performed on Cr$_{100 ML}$/W(110), was reported. On the other hand, particularly 
PE provides mostly direct insight into the structure of the occupied electron states allowing 
better understanding of the SDW in solids. The incommensurability of the SDW in Cr can cause (i) 
corrections of the ground-state electronic structure mainly in the region of the Fermi-energy 
(E$_F$) gap(s) related to the magnetic order, and (ii) renormalization of the electronic bands due 
to the electron interaction with the accompanying SDW state excitations similar to the case of 
quasiparticle interactions in, e.g., high-temperature superconductors \cite{Shen}. Both phenomena 
can be studied with angle-resolved PE. The effect of SDW-derived folding of the antiferromagnetic 
PE bands is expected to be negligible \cite{Mol} by the reason of only weak perturbation of the 
crystal potential introduced by the large unit cell of the incommensurate-SDW chromium. 
 
In the present paper we approach understanding of the SDW phenomena in multilayer Cr systems by an 
angle-resolved PE study of epitaxial Cr films (10 to 100 ML) grown on W(110). The data are 
compared with the results of local spin density approximation (LSDA) layer-Korringa-Kohn-Rostoker 
(LKKR) calculations. It is shown that the way of characterization of the SDW state by looking for 
the Fermi-energy gap \cite{Schaf}, is not straightforward, since the gap is mostly located above 
E$_F$ and masked by surface states. Instead, the incommensurate SDW can be monitored and important 
parameters of the SDW-related interactions, such as coupling strength and energy of magnetic 
excitations, can be determined from the dispersion of the electronic bands close to E$_F$. 

\begin{figure}
\includegraphics[width=86mm,angle=0,clip]{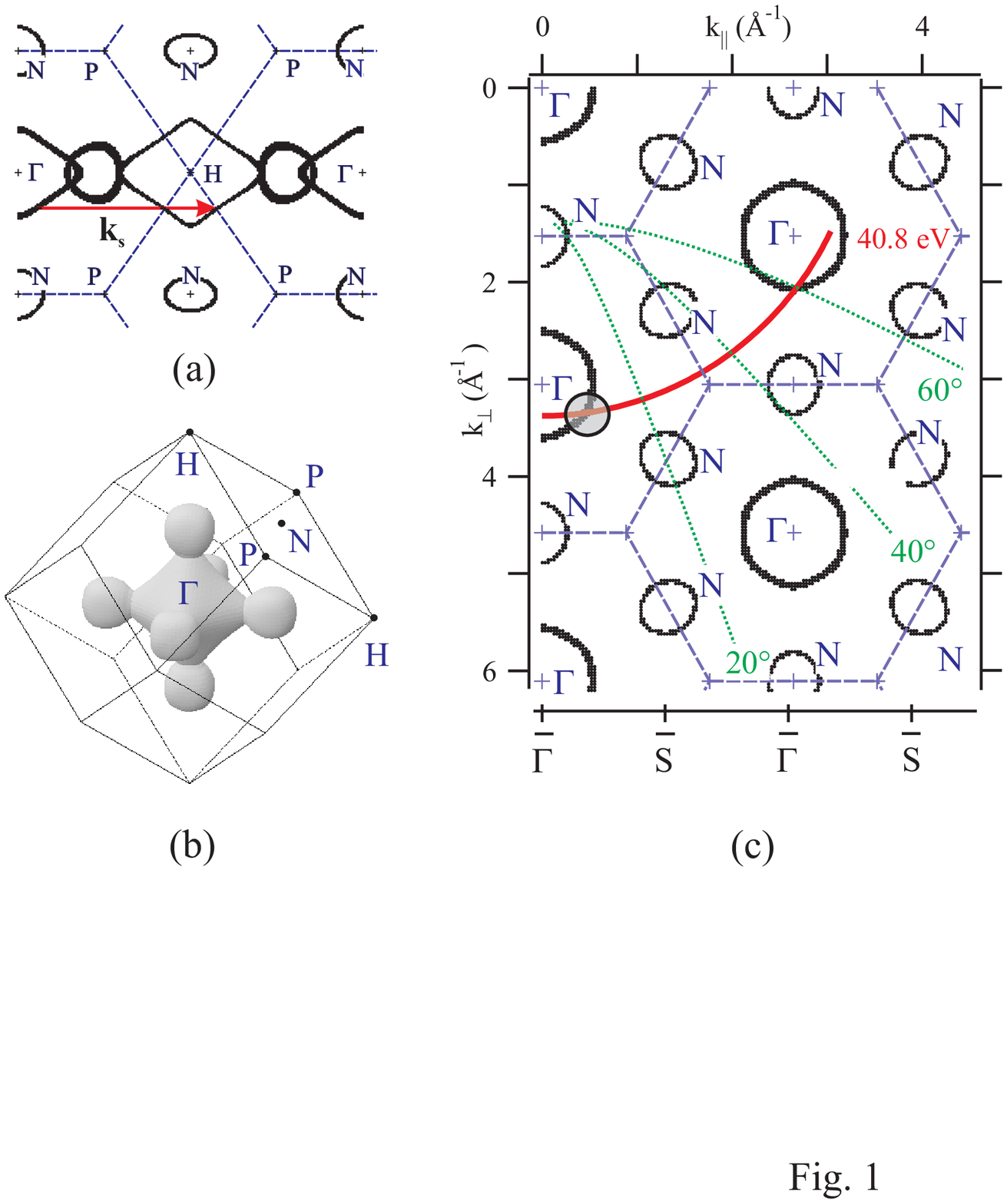}  
\caption{ (a) FS cut within the (110) plane of the bulk BZ of $bcc$ Cr. (b) FS jack around the 
$\Gamma$ point. (c) Path in the bulk BZ of $bcc$ Cr sampled with h$\nu$ = 40.8 eV (thick line). 
${\bf k}_{\parallel}$ and ${\bf k_{\bot}}$  denote parallel and perpendicular to the (110) plane 
components of the wave vector, respectively.\label{fig1} }
\end{figure}

Films of Cr were grown {\it in situ} on a W(110) crystal by deposition from a molybdenum crucible 
heated by electron beam. Various thicknesses of Cr were used in order to follow the transition 
from the thick films characterized by the SDW state (47 and 100 ML) to the thin film (10 ML), 
where the incommensurate SDW is suppressed \cite{Nikl}. ''As deposited'' samples were annealed to 
$900^{\circ}$C in order to ensure well-ordered films under consideration. In all cases the grown 
epitaxial films revealed sharp low-energy electron diffraction patterns. The PE measurements were 
performed with a SCIENTA 200 electron-energy analyzer using monochromatized light from a He lamp 
(h$\nu$ = 40.8 eV). The overall-system energy resolution was set to 130 meV full width at half 
maximum, the angular resolution to $0.4^{\circ}$. All experiments were carried out at room 
temperature well below T$_N$ at the surface of Cr(110) \cite{Schaf}, the base pressure was 
$6\times10^{-9}$ Pa. 

The electronic structure of the Cr(110) semi-infinite crystal was calculated within a LSDA-LKKR 
multiple scattering approach. This method uses the Green's function technique and allows the 
calculation of systems with a broken translational symmetry along one direction \cite{halil}. The 
calculated layer resolved spectral density of states are related to the layer Green's function 
simply as $D(\mathbf{k}_{\parallel},E) = -\mathrm{Tr\,Im}\,G(\mathbf{k}_{\parallel},E)/\pi$.

\begin{figure}[b!]
\includegraphics[width=86mm,angle=0,clip]{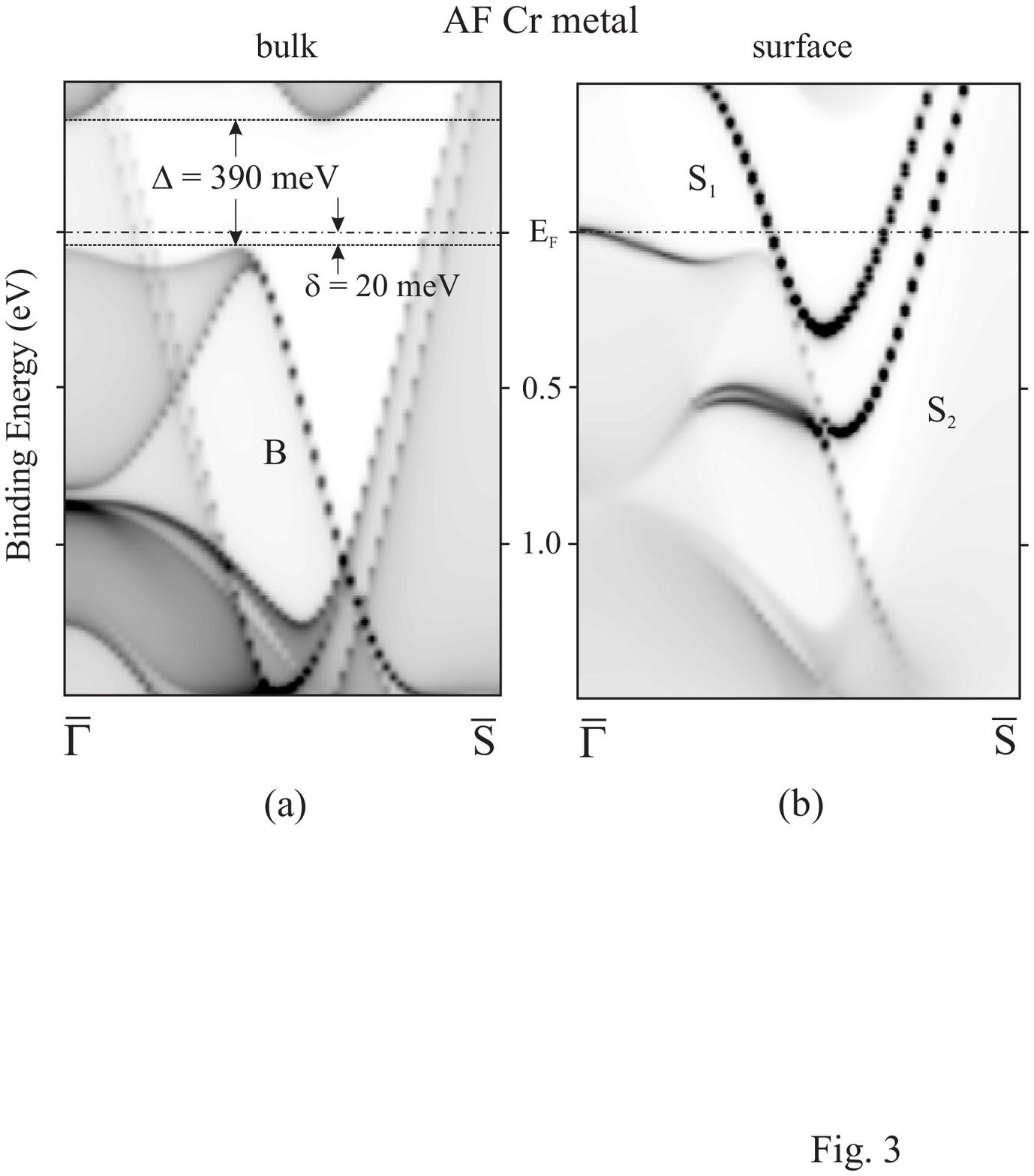}  
\caption{ Spectral LKKR-DOS. 
    $\mathbf{k}_{\parallel}$ along the $\overline{\Gamma}-\overline{S}$ direction.
    The darker the color, the more intense the DOS.\label{fig2} }
\end{figure}

A calculated cut of the bulk FS of $bcc$ Cr within the (110) plane is demonstrated in Fig. 1(a). 
The rhomb-like contours of the FS around the $\Gamma$ and the H points of the bulk BZ look almost 
identical: They are connected by the spanning vector ${\bf k}_s$ and are expected to be strongly 
affected by the magnetic ordering. Also the three-dimensional FS jacks at the $\Gamma$ and the H 
points [the $\Gamma$-point jack is shown in Fig. 1(b)] have similar shapes. In a first 
approximation they can be obtained from each other by a parallel transfer defined by ${\bf k}_s$. 
Therefore, everywhere in the region of these jacks one would expect the discussed above effects of 
the energy gap and the band renormalization related to the SDW state. We have performed 
experiments along the $\overline{\Gamma}-\overline{S}$ direction in the surface BZ of Cr by 
varying the polar electron-emission angle. In this way the part of the FS around the $\Gamma$ 
point, where the bumps at the corners [see Fig. 1(b)] do not distort the measurements, was 
sampled. Assuming free-electron like final states, the measurements were carried out along the 
path in the bulk BZ as shown in Fig. 1(c), where for simplicity the FS calculated for $bcc$ Cr is 
presented. The BZ for AF phase can be obtained from the $bcc$ BZ by folding. The path crosses the 
FS sheet in the region of interest marked by a shaded circle in the figure. 

The results of the LKKR spectral density of states (DOS) calculations for the bulk and the surface 
layer of AF $sc$ Cr without the incommensurate SDW modulation are presented in Fig. 2. Similar to 
other theoretical data \cite{Schaf,Saki,Falic} the AF energy gap is obtained in the ${\bf 
k}_{\parallel}$ region marked in Fig. 1(c). Band B that crosses E$_F$ in the paramagnetic phase is 
now folded back toward low binding energies (BEs) not reaching the Fermi energy (Fig. 2). As a 
result an indirect AF gap $\Delta$ of about 390 meV is formed. Note that the calculated gap is 
predominantly located in the region of the unoccupied electron states.

The corresponding ''as measured'' data close to the energy-gap region taken for the 100-ML thick 
Cr layer are shown in a gray-scale plot in Fig. 3(a). The measured band follows the behavior of 
the calculated band B. As ${\bf k}_{\parallel}$ increases it approaches first the Fermi energy. At 
${\bf k}_{\parallel}$ $>$ 0.55 \AA$^{-1}$ it turns back toward higher BEs. The PE intensity of the 
band in this ${\bf k}_{\parallel}$ region is almost negligible, a cross-section effect, which is 
well known for folded bands in solids \cite{Mol}. From Fig. 3(a) and further careful analysis of 
the individual PE spectra, it is, however, by far not clear as to whether the gap is seen in the 
E$_F$ region. On the other hand, according to our calculations only small part of the AF gap 
($\delta \sim$ 20 meV, Fig. 2) is found below E$_F$ and can directly be observed with PE. To 
extract information about unoccupied electron states we followed the method used in Refs. 
\cite{Schaf,Rein}. The raw data were divided by the Fermi distribution to allow observation of the 
thermally excited states. The width of 
\begin{figure}[b!]
\includegraphics[width=85mm,angle=0,clip]{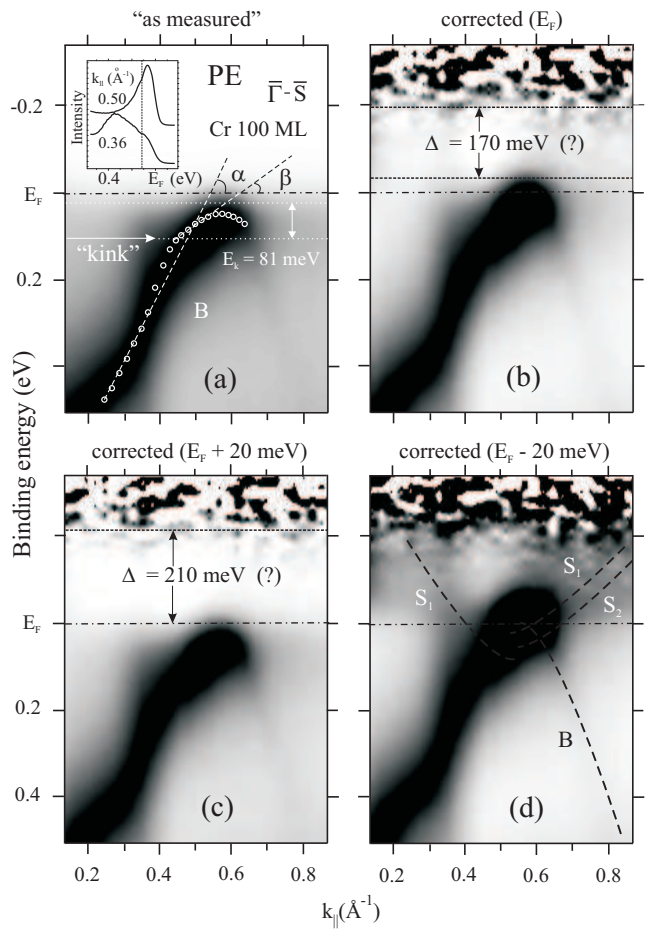}  
\caption{ Logarithmic PE signal. Dark areas represent higher intensity. (a) ''As measured'' data. 
Energy of the band B maximum for each PE spectrum are marked with white circles. Inset: two 
spectra taken below (${\bf k}_{\parallel}$ = 0.36 \AA$^{-1}$) and above (0.50 \AA$^{-1}$) the kink 
position. Vertical line is the kink energy. (b) Data corrected by the Fermi distribution with 
$\mu$ equal to the measured E$_F$. In (c) and (d) $\mu - E_F$ was selected to be $\pm20$ meV, 
respectively.\label{fig3} }
\end{figure}
the Fermi distribution, which is about 100 meV at room temperature, gives a scale for the 
accessible energies. The corrected data are presented in Fig. 3(b). It seems that indeed an energy 
gap in the region of the unoccupied states of $\sim$ 170 meV is monitored here. Note, however, 
that the corrected results depend strongly on the used value of the chemical potential $\mu$ that 
was estimated from the measurements of E$_F$ for a metallic sample. Qualitatively different 
results are obtained shifting $\mu$ by only 20 meV toward higher or lower BEs, variations, which 
are much smaller than the energy resolution of the experiments. An increase of the chemical 
potential leads to a considerably larger value of the derived energy gap [Fig. 3(c)]. Even more 
drastic changes are observed upon decrease of $\mu$ [Fig. 3(d)]. The AF gap is not monitored 
anymore. Instead, the region between the Fermi energy and 0.2 eV above E$_F$ is filled with 
electron states. The surface origin of these states is seen from a comparison with the theoretical 
results shown in the right panel in Fig. 2.

One has to admit that the proposed method to use the Fermi-distribution correction of the PE data 
\cite{Schaf,Rein} is not straightforward even to extract the energy-gap information related with 
the antiferromagnetism, which is the main contribution into the magnetic order in Cr metal below 
T$_N$. In this respect, it seems there is no way to follow fine gap changes that might be caused 
by the rather weak incommensurate SDW contribution. In difference to the energy gap the 
SDW-derived renormalization of the shallow electronic bands in Cr systems can easily be monitored. 
As seen in Fig. 2, the calculated band B reveals smooth monotonic dispersion when going from the 
$\overline{\Gamma}$ point toward the gap at E$_F$. In contrast to that our experimental band shows 
a pronounced ''kink'' at ${\bf k}_{\parallel}$ $\sim$ 0.45 \AA$^{-1}$. The observed kink can be 
simulated by a superposition of two contributions: the calculated single-particle LSDA band for 
small ${\bf k}_{\parallel}$ and a renormalized one for ${\bf k}_{\parallel}$ $>$ 0.45 \AA$^{-1}$. 
The PE spectra of band B in the region of the kink [inset in Fig. 3(a)] have an asymmetric 
lineshape with two structures: a main peak and a shoulder, which exchange their spectral weight 
crossing the kink. The main peak, which is quit broad at high BE, becomes much sharper close to 
E$_F$. All above evidences that the signals measured in the dispersion region before and after the 
kink have different nature.

\begin{figure}
\includegraphics[width=86mm,angle=0,clip]{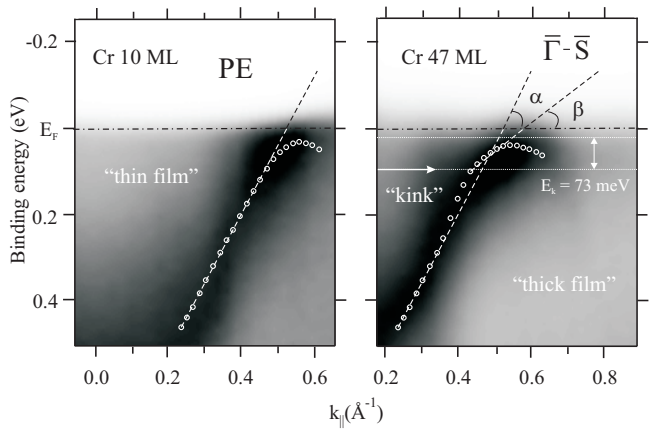}  
\caption{ PE signal in the region of the kink.\label{fig4} }
\end{figure}

This kind of behavior of bands close to E$_F$ is a well known phenomenon for correlated systems 
\cite{Shen}. There is a pole in the self-energy $\Sigma$ of the material at the energy of a 
quasiparticle interaction. The strength of interaction is described via a coupling constant 
$\lambda$ that is defined as $\lambda = - \delta Re\Sigma/\delta E|_{E_F}$. The energy dependence 
of $\Sigma$ can be inverted into a {\bf k} dependence. Thereby (i) the self-energy pole is 
transformed into the kink in the band dispersion and (ii) the coupling constant is rewritten as 
$\lambda = [(\delta E^{LSDA}/\delta {\bf k})/(\delta E^{ren}/\delta {\bf k})|_{E_F}-1]$ depending 
on the ratio of the group velocities determined by the LSDA [$E^{LSDA}({\bf k})$] and the 
renormalized [$E^{ren}({\bf k})$] bands. As a result both $\lambda$ and the energy of 
quasiparticle excitation can be derived from the analysis of the band dispersion in the vicinity 
of the kink. Energy gaps that can appear to stabilize the excited state may complicate the 
situation. In this case the quasiparticle energy is related not to E$_F$, but to the bottom of the 
corresponding gap.

For the 100-ML film the kink energy relative to the bottom of the calculated Fermi-energy gap is 
estimated to be (81$\pm$7) meV [Fig. 3(a)] that is of the order of the expected energy for the 
magnon excitations accompanying the incommensurate SDW in $sc$ Cr metal \cite{Fawc,Lui}. 
Interaction with phonons can be ruled out by the reason of lower energy of the phonon excitations 
\cite{Fawc}. To obtain $\lambda$ the ratio of the group velocities for the LSDA and the 
renormalized bands was substituted by the ratio of the tangents of two angles $\alpha$ and $\beta$ 
between the directions of the corresponding band dispersion and the ${\bf k}_{\parallel}$ axis 
[see Fig. 3(a)]. The directions of band dispersion were approximated by straight lines through the 
energy positions of the main peak of each individual spectrum that are shown by white circles in 
the figure. By this procedure a value $\lambda = 1.41\pm0.09$ was obtained pointing to a moderate 
strong quasiparticle interaction.     

The behavior of band B in the region of the kink was studied for Cr films of different thicknesses 
(Fig. 4). In all cases the films were selected to be thick enough to reveal both the bulk and the 
surface features of the electronic structure of the AF $sc$ Cr (Fig. 2). It is of high importance 
to underline that no kink is observed for the 10-ML film, where the incommensurate SDW is 
suppressed \cite{Nikl}. This fact is considered as a strong evidence for the SDW in thicker Cr 
films and the possibility to monitor this state in the dispersion of the renormalized bands. A 
decrease of the film thickness from 100 to 47 ML does not result in the removal of the kink, 
although it causes a slight drop of its energy to (73$\pm$7) meV. Assuming a linear energy 
dispersion of the spin-wave excitation \cite{Fawc} this may be assigned to the increase of the 
bulk period of the related SDW due to growing influence of the boundary conditions. Also $\lambda$ 
decreases slightly with the thickness reaching the value $1.30\pm0.09$ for the 47-ML thick film. 
As reported in Ref. \cite{Urg} there are two phase slips in the coupling between the marginal 
layers in the range from 47- to 100-ML thick Cr spacers in Fe/Cr/Fe(100) systems. According to 
Niklasson {\it et al.} \cite{Nikl} each slip originates in a jump from one branch of the SDW to 
another. Therefore, the observed change of $\lambda$ may be understood by slightly different 
electron interaction with magnons associated with individual SDW branches, which are expected to 
be also present in Cr/W(110).  

{\it In summary} we have shown that angle-resolved PE can successfully be used to monitor the SDW 
in thin films of Cr. A valuable information about quasiparticle magnetic interactions was obtained 
by the analysis of the dispersion of the renormalized electronic bands in the vicinity of E$_F$. 
The used approach can be applied to a large variety of other SDW systems including magnetic 
multilayer structures highly relevant for technological applications.   

This work was supported by the DFG, SFB 463 TP B16. The authors are grateful to M. Richter, S. 
Halilov, C. Laubschat, S.-L. Drechsler, S. Shulga, S. Borisenko, A. Kordyuk, and T. Kim  for 
valuable scientific discussions.

\end{document}